\documentclass{llncs}

\usepackage{microtype}%if unwanted, comment out or use option "draft"
\usepackage{amsmath}
\usepackage{graphicx}
\usepackage[boxed]{algorithm2e}
\usepackage{multirow}
\DeclareGraphicsExtensions{.pdf,.png}
\usepackage{xcolor}
\usepackage{tikz}
\usepackage[linguistics]{forest}

\title{Faster Dynamic Compressed $d$-ary Relations%
\thanks{Funded by the Millennium Institute for Foundational Research on Data
(IMFD), Chile and by European Union's Horizon 2020 research and innovation
programme under the Marie Sklodowska-Curie grant agreement No 690941 (project
BIRDS). GdB funded by Xunta de Galicia/FEDER-UE CSI:ED431G/01 and GRC:ED431C
2017/58; by MINECO-AEI/FEDER-UE TIN2016-77158-C4-3-R; and by MICINN RTC-2017-5908-7.}}

\author{Diego Arroyuelo \inst{1,2}
\and
Guillermo de Bernardo \inst{3}
\and
Travis Gagie \inst{4}
\and 
{ \\ Gonzalo Navarro \inst{1,5}}
}

\institute{
Millennium Institute for Foundational Research on Data (IMFD), Chile
\and
Departamento de Inform\'atica, Universidad T\'ecnica Federico Santa Mar\'{\i}a,
Chile, \texttt{darroyue@inf.utfsm.cl}
\and
University of A Coru\~na, A Coru\~na, Spain, \texttt{gdebernardo@udc.es}
\and
Faculty of Computer Science, Dalhousie University, Halifax, Canada,
\texttt{travis.gagie@gmail.com}
\and
Department of Computer Science, University of Chile, Chile, \texttt{gnavarro@dcc.uchile.cl}}

\begin{document}

\maketitle

\begin{abstract} 
The $k^2$-tree is a successful compact representation of binary relations that
exhibit sparseness and/or clustering properties. It can be extended to $d$
dimensions, where it is called a $k^d$-tree. The representation boils 
down to a long
bitvector. We show that interpreting the $k^d$-tree as a dynamic trie on the 
Morton codes of the points, instead of as a dynamic representation of the 
bitvector as done in previous work, yields operation times that are below the
lower bound of dynamic bitvectors and offers improved time performance in 
practice.
\end{abstract}

\section{Introduction}

The $k^2$-tree \cite{BLN13} is a compact data structure conceived to represent
the adjacency matrix of Web graphs, but its functionality was later extended 
to represent other kinds of $d$-ary relations such as ternary relations
\cite{AGdBBN16}, point grids \cite{BdBKNS16}, raster data \cite{dBABNP13}, 
RDF stores \cite{AGBFMPN14}, temporal graphs \cite{CdBFPS18}, graph databases
\cite{AGFLP18}, etc.

The $k^2$-tree compactly represents an extension of a variant of the Quadtree
data structure \cite{Mor66}, more precisely of the MX-Quadtree \cite[Section
1.4.2.1]{Sam06}. The MX-Quadtree splits the $n \times n$ grid into four
submatrices of $n/2 \times n/2$. The root indicates which of the submatrices
are nonempty of points, and a child of the root recursively represents each
nonempty submatrix. In the $k^2$-tree, the matrix is instead
split into $k^2$ submatrices of $n/k \times n/k$ cells. In $d$ dimensions,
the structure becomes a $k^d$-tree, where the grid is
divided into $k^d$ submatrices of $n/k \times \cdots \times n/k$ cells.
The height of the $k^d$ tree is then $\log_{k^d} (n^d) = \log_k n$.

Instead of using pointers to represent the tree topology, the $k^d$-tree uses
a long bitvector $B[1..N]$, where each node stores only $k^d$ bits indicating 
which of its submatrices are nonempty, and all the node bitvectors are 
concatenated level-wise into $B$. Bitvector $B$ supports navigation towards
children and parents in $O(1)$ time \cite{BLN13} by means of rank/select 
operations \cite{Cla96,Mun96} on bitvector $B$. Query operations like
retrieving all the neighbors or the reverse neighbors of a node (when
representing graphs) or retrieving all the points in a range (when
representing grids) then translate into traversals on the
$k^d$-tree \cite{BLN13}. 

In various applications one would like the relations to be {\em dynamic},
however, that is, elements (graph edges, grid points) can be inserted and 
deleted from the relation. Each such update requires flipping bits or 
inserting/deleting chunks of $k^d$ bits at each of the $\log_k n$ levels in $B$.
Such operations can be supported using a dynamic bitvector representation 
\cite{BCPdBN17}. There exists, however, 
an $\Omega(\log N/\log\log N)$ lower bound to support updates and rank/select 
operations on a bitvector of length $N$ \cite{FS89}, and such slowdown factor 
multiplies every single operation carried out on the bitvector, both for 
traversals and for updates. 

In this paper we take a different view of the $k^d$-tree representation. We
regard the $k^d$-ary tree as a trie on the Morton codes \cite{Mor66} of the
elements stored in the grid. The Morton code (in two dimensions, but the
extension is immediate) is the concatenation of the $\log_k n$ 
identifiers of the consecutive subgrids chosen by a point until
it is inserted at the last level. We then handle a trie of strings of length
$\log_k n$ over an alphabet of size $k^d$. While such a view yields no
advantage in the static case, it provides
more efficient implementations in the dynamic scenario. For example, a
succinct dynamic trie \cite{ADR16} on the Morton codes requires
space similar to our bitvector representation, but it is much faster
in supporting the operations: $o(d\log k)$ time, and constant for practical
values of $d$ and $k$.

In this paper we implement this idea and show that it is not only
theoretically appealing but also competitive in practice with the preceding
dynamic-bitvector-based representation \cite{BCPdBN17}. In our way, we define
a new depth-first deployment for tries that, unlike the level-wise one
\cite{BLN13}, cannot be traversed in constant time per edge. Yet, we show it 
turns out to be convenient in a dynamic scenario because we have to scan only
small parts of the representation.

\section{The $k^2$-tree and its representation as a trie}

Let us focus on the case $k=2$ and $d=2$ for simplicity; $d=2$ encompasses all
the applications where we represent graphs, and the small value of $k$ is the
most practical in many cases. Given $p$ points in an $n \times n$ matrix $M$, 
the $k^2$-tree is a $k^2$-ary (i.e., $4$-ary) tree where each node represents 
a submatrix. Assume $n$ is a power of $k$ (i.e., of $2$) for simplicity. The 
root then represents the whole 
matrix $M[0..n-1,0..n-1]$. Given a node representing a submatrix
$M[r_1..r_2,c_1..c_2]$, its $4$ children represent the submatrices
$M[r_1..r_m,c_1..c_m]$ (top-left), $M[r_1..r_m,c_m+1..c_2]$ (top-right), 
$M[r_m+1..r_2,c_1..c_m]$ (bottom-left), and $M[r_m+1..r_2,c_m+1..c_2]$
(bottom-right), in that order, where $r_m = (r_1+r_2-1)/2$
and $c_m = (c_1+c_2-1)/2$. Each of the $4$ submatrices of a node may be empty 
of points, in which case the node does not have the corresponding child.
The node stores $4$ bits indicating with a $1$ that the corresponding matrix
is nonempty, or with a $0$ that it is empty.
The $k^2$-tree is of height $\log_k n = \log_2 n$. See Figure~\ref{fig:k2tree}.

\begin{figure}[t]
 \centering

  \begin{tikzpicture}[scale=.25]
    \draw[red,line width=0.5pt] (0,0) grid[step=1] (16,16);
    \draw[black,line width=1pt]   (0,0) grid[step=2] (16,16);
    \draw[red,line width=1.5pt] (0,0) grid[step=4] (16,16);
    \draw[black,line width=2pt] (0,0) grid[step=8] (16,16);
    \draw[black,line width=2pt] (0,0) rectangle (16,16);

    \fill (2.5, 15.5) circle (0.35cm);
    \fill (3.5, 15.5) circle (0.35cm);
    \fill (4.5, 15.5) circle (0.35cm);
    \fill (5.5, 15.5) circle (0.35cm);
    \fill (6.5, 15.5) circle (0.35cm);

    \fill (7.5, 14.5) circle (0.35cm);
    \fill (3.5, 14.5) circle (0.35cm);

    \fill (1.5, 13.5) circle (0.35cm);

    \fill (0.5, 11.5) circle (0.35cm);
    \fill (1.5, 11.5) circle (0.35cm);

    \fill (3.5, 8.5) circle (0.35cm);

    \fill (12.5, 7.5) circle (0.35cm);

    \fill (12.5, 4.5) circle (0.35cm);
    
  \end{tikzpicture}

\vspace{2mm}

\begin{forest}for tree={inner sep=0pt,outer sep=-1pt}
  [~~1$^\textbf{(\textcolor{red}{0})}$
    [~~1$^\textbf{(\textcolor{red}{1})}$
      [~~1$^\textbf{(\textcolor{red}{3})}$
        [0] 
        [~~1$^\textbf{(\textcolor{red}{7})}$
         [1][1][0][1]
        ]
        [~~1$^\textbf{(\textcolor{red}{8})}$
         [0][1][0][0]
        ]
        [0]
      ] 
      [~~1$^\textbf{(\textcolor{red}{4})}$
        [~~1$^\textbf{(\textcolor{red}{9})}$
         [1][1][0][0]
        ]
        [~~~~1$^\textbf{(\textcolor{red}{10})}$
         [1][0][0][1]
        ]
        [0]
        [0]
      ]
      [~~1$^\textbf{(\textcolor{red}{5})}$
       [~~~~1$^\textbf{(\textcolor{red}{11})}$
        [1][1][0][0]
       ]
       [0]
       [0]
       [~~~~1$^\textbf{(\textcolor{red}{12})}$
        [0][0][0][1]
       ]
      ]
      [~~~~0~~~~]
    ]
    [~~~~~~~~~~0~~~~~~~~~~~
    ]
    [~~~~~~~~~0~~~~~~~~~
    ]
  [~~1$^\textbf{(\textcolor{red}{2})}$
    [0]
    [~~1$^\textbf{(\textcolor{red}{6})}$
        [~~1$^\textbf{(\textcolor{red}{13})}$
          [1][0][0][0]
        ]
        [0]
        [~~1$^\textbf{(\textcolor{red}{14})}$
          [0][0][1][0]
        ]
        [0]
    ]
    [0]
    [0]
 ]
]
\end{forest}

\vspace{2mm}

    \begin{tabular}{c|c|c|c|c|c|c|c|c|c|c|c|c|c|c}
    \hline
    \textcolor{red}{0} & \textcolor{red}{1} & \textcolor{red}{2} & \textcolor{red}{3} & \textcolor{red}{4} & \textcolor{red}{5} &\textcolor{red}{6} & \textcolor{red}{7} & \textcolor{red}{8} & \textcolor{red}{9} & \textcolor{red}{10} & \textcolor{red}{11} & \textcolor{red}{12} & \textcolor{red}{13} & \textcolor{red}{14} \\
        
    \texttt{1001} & \texttt{1110} & \texttt{0100} & \texttt{0110} & \texttt{1100} & \texttt{1001} & \texttt{1010} & \texttt{1101} & \texttt{0100} & \texttt{1100} & \texttt{1001} & \texttt{1100} & \texttt{0001} & \texttt{1000} & \texttt{0010}\\
    \end{tabular}

\caption{Binary relation for the set $\{(0,2)$, $(0,3)$, $(0,4)$, $(0,5)$, $(0,6)$, $(1,3)$, $(1,7)$, $(2,1)$, $(4,0)$, $(4,1)$, $(7,3)$, $(8,13)$, $(11,13)\}$ (on top). The corresponding $k^2$-tree (in the middle), and its levelwise representation (on the bottom).}
\label{fig:k2tree}
\end{figure}

\paragraph{Succinct representation.}

A simplified description of the compact $k^2$-tree representation \cite{BLN13} 
consists of a bitvector $B$ where the tree is traversed levelwise, left to 
right, and the $k^2=4$ bits of all the nodes are concatenated. Then, if the 
tree has $v$ nodes, the
bitvector $B$ is of length $k^2v = 4v$, $B[1..4v]$. Note that the nodes of depth
$\log_k n = \log_2 n$ correspond to $4$ cells, and therefore it is sufficient 
to store their $4$ bits; their children are not represented. Given $p$ points, 
the number of nodes of the $k^2$-tree is $v \le p\log_4(n^2/p)+O(p)$ 
\cite[Sec.~9.2]{Nav16}.

Each $k^2$-tree node is identified by the position of the
first of the $4$ bits that describes
its empty/nonempty children. To move from a node $i$ to its $t$-th child, the
formula is simply $4 \cdot rank_1(B,i)+t$, where $rank_1(B,i)$ counts the
number of 1s in $B[1..i]$ and can be computed in $O(1)$ time using $o(v)$
space on top of $B$ \cite{Cla96}. For example, we determine in $O(\log_k n)$
time whether a certain point exists in the grid. Other operations require
traversal of selected subtrees \cite{BLN13}.

\paragraph{Dynamic $k^2$-trees.}

A dynamic $k^2$-tree \cite{BCPdBN17} is obtained by representing $B$ as a
dynamic bitvector. Now operation $rank$ takes time $O(\log v/\log\log v)$
\cite{NS14},
which is optimal \cite{FS89}. This slows down the structure with respect to 
the static variant. For example, determining whether a point exists takes time 
$O(\log_k n \cdot \log v/\log\log v) \subseteq O(\log^2 n /\log\log n)$.
To insert a point $(r,c)$, we must create its path up to the leaves, converting
the first $0$ in the path to a $1$ and thereafter inserting groups of $k^2=4$
bits, one per level up to level $\log_k n$. This takes time
$O(\log_k n \cdot \log v/\log\log v)$ as well. Deleting a point is analogous.

\paragraph{Morton codes.}

Consider a point $(r,c)$, which induces a root-to-leaf path in the $k^2$-tree. 
If we number the $4$ submatrices described in the beginning of this section as
0,1,2,3, then we can identify $(r,c)$ with a sequence of $\log_4 (n^2) = 
\log_2 n$ symbols
over the alphabet $[0..3]$ that indicate the submatrix chosen by $(r,c)$ at
each level. In particular, note that if we write the symbols in binary,
$0=00$, $1=01$, $2=10$, and $3=11$, then the row $r$ is obtained by
concatenating the first bits of the $\log_2 n$ levels, from highest to lowest
bit, and the column $c$ is obtained by concatenating the second bits of the
$\log_2 n$ levels. The Morton code of $(r,c)$ is then obtained by interlacing
the bits of the binary representations of $r$ and $c$. 

As a consequence, we can regard the $k^2$-tree as the trie of
the Morton codes of all the $p$ points, that is, a trie storing $p$ strings of
length $\log_k n = \log_2 n$ over an alphabet of size $k^2=4$.
The extension to general values of $k^d$ is immediate.

\paragraph{Succinct tries.} A recent dynamic representation \cite{ADR16} of 
tries of $v$ nodes over alphabet $[0..\sigma-1]$ requires $v(2+\log_2 \sigma) +
o(v\log\sigma)$ bits. If $\sigma$ is polylogarithmic in $v$, it simulates each 
step of a trie traversal in $O(1)$ time, and the insertion and deletion of
each trie node in $O(1)$ amortized time. 
Used on our Morton codes, with alphabet size
$\sigma=k^2=4$, the tries use $v(2+2\log_2 k) + o(v) = 4v+o(v)$ bits, exactly
as the representation using the bitvector $B$. Instead, they support
queries like whether a given point exists in time $O(\log_k n)$,
and inserting or deleting a point in amortized time $O(\log_k n)$, way faster
than on the dynamic bitvector $B$.

\paragraph{The general case.}
With larger values of $k$ and $d$, $B$ requires $k^d v$ bits, and it may become 
sparse. By using sparse bitvector representations \cite{OS07}, the space 
becomes $O(p\log(n^d/p) + pd\log k)$ bits \cite[Sec.~9.2]{Nav16}, but the time 
of operation {\em rank} becomes $O(d\log k)$, and this time penalty factor 
multiplies all the other operations. A dynamic representation of the
compressed bitvector \cite{NS14} uses the same space and requires $O(\log
v/\log\log v)$ time for each operation.
The space usage of the trie \cite{ADR16} on a general alphabet of size 
$\sigma=k^d$ is of the same order, $O(p\log(n^d/p) + pd\log k)$ bits,
but the operations are supported in less time,
$O(\log\sigma/\log\log\sigma) = 
O(\log(k^d)/\log\log(k^d))=O(d\log k/ \log(d\log k))$ (amortized for updates).
The insertion or deletion of a point, which affects $\log_k n$ tree edges,
then requires $O(d\log n / \log(d\log k))$ amortized time.
We state this simple result as a theorem.

\begin{theorem} \label{thm:main}
A dynamic $k^d$ tree can represent $p$ points on a $n^d$-size grid within
$O(p\log(n^d/p)+pd\log k)$ bits, while supporting the traversal, insertion,
or deletion of each tree edge in time $O(d\log k/\log(d\log k))$ (amortized for
updates). If $k^d = O(\mathrm{polylog}\,p)$, then the times are $O(1)$
(also amortized for updates).
\end{theorem}

\section{Implementation of the dynamic trie}

We now define a practical implementation of succinct dynamic tries, for the particular case of $k^2$-trees with $k=2$. The whole trie is divided into blocks, each being a connected component of the trie. A block can have child blocks, so we can say that the trie is represented as a tree of blocks. Let us define values $N_1 < N_2 < \cdots < N_{max}$, such that $N_i = N_{i-1}/\alpha$, for $i = 2,\ldots, max$, for a given parameter $0 < \alpha < 1$, and $N_{max}=4\cdot N_1$ \cite{AN11}. At any given time, a block $B$ of size $N_i$ is able to store at most $N_i$ nodes. If new nodes are added to $B$ such that the number of nodes exceeds $N_i$, then $B$ is grown to have size $N_j$, for $j > i$, such that the new nodes can be stored. By defining the block sizes as we do, we ensure that the fill ratio of each block is at least $1-\alpha$; for example, if $\alpha=0.05$, then every node is at least $95\%$ full, which means that the space wasted is at most $5\%$. 

Each block $B$ stores the following components:
\begin{itemize}
    \item $T_{\!B}$: the tree topology of the connected component represented by the block. Every node in the trie is either an internal node, a leaf node, or 
a {\em frontier} node in some $T_{\!B}$. The latter are seen as leaves
in $T_{\!B}$, but they correspond to trie nodes whose subtree is stored in a descendant block. We mark such nodes in $B$ and store a pointer to the corresponding child block, see next.
    \item $F_{\!B}$: a sorted array storing the preorder numbers of the frontier nodes. 
    \item $P_{\!B}$: an array with the pointers to children blocks, in the same order of $F_{\!B}$. 
    \item $d_{\!B}$: the depth (in the trie) of the root of $T_{\!B}$.
\end{itemize}

Unlike the classical $k^2$-tree representation \cite{BLN13,BCPdBN17}, which 
deploys the nodes levelwise, we represent the tree topology $T_{\!B}$ in 
depth-first order. This order is compatible with our block layout and speeds 
up the insertion and deletion of points, since the bits of all the edges
to insert or remove are contiguous.

\paragraph{Representation.}
In $T_B$, each node is encoded using 4 bits, indicating which of its children are present. For instance, `\texttt{0110}' encodes a node that has two children, labeled by symbols \texttt{1} and \texttt{2}. Therefore, the total number of bits used to encode the trees $T_B$ is {\em exactly the same} as in the classical representations \cite{BLN13,BCPdBN17}.

\begin{figure}[t]
    \centering
    \includegraphics[width=0.9\textwidth]{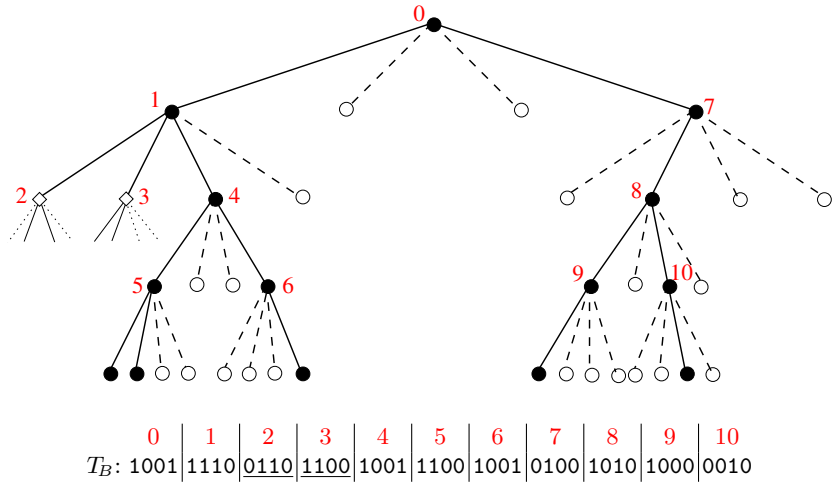}
    
    \vspace{0.5cm}
    
    \begin{tabular}{cc|c|c|c|c|c|c|c|c|c|c}
       & \textcolor{red}{0} & \textcolor{red}{1} & \textcolor{red}{2} & \textcolor{red}{3} & \textcolor{red}{4} & \textcolor{red}{5} &\textcolor{red}{6} & \textcolor{red}{7} & \textcolor{red}{8} & \textcolor{red}{9} & \textcolor{red}{10} \\
        
    $T_{\!B}$: &  \texttt{1001} & \texttt{1110} & \texttt{\underline{0110}} & \texttt{\underline{1100}} & \texttt{1001} & \texttt{1100} & \texttt{1001} & \texttt{0100} & \texttt{1010} & \texttt{1000} & \texttt{0010}\\
    \end{tabular}
    
    \caption{Example block of a $k^2$-tree and its depth-first representation. Depth-first numbers are shown along with each node, and they correspond with the index in the array representation. Nodes with numbers 2 and 3 (underlined in $T_B$) are frontier nodes.}
    \label{fig:k2treedyn}
\end{figure}

We store $T_{\!B}$ using a simple array able to hold $N_i$ nodes. A node is identified by its index within this array. Figure~\ref{fig:k2treedyn} shows an example top block for the $k^2$-tree of Figure~\ref{fig:k2tree} and our array-based depth-first representation. Depth-first numbers are shown along each node; these are also their indexes in the array storing $T_B$. In the example, nodes with depth-first number 2 and 3 are frontier nodes; they are underlined in the array representation.

Apart from $T_B$,
each block $B$ then requires $3$ words to store $d_B$ and its corresponding 
entries in the arrays $F_{B'}$ and $P_{B'}$ in its parent block $B'$. This
implies a maximum overhead of $O(\log(n)/N_1)$ bits per node, assuming pointers
of $\Theta(\log n)$ bits as in the transdichotomous RAM model of computation.
Thus we have to choose $N_1 = \omega(\log N)$ for this overhead to be $o(n)$.

The depth-first order we use, however, corresponds more to the \textsc{dfuds} representation \cite{BDMRRR05}, whereas the classical levelwise deployment is analogous to a \textsc{louds} representation \cite{Jac89}. An important difference is that, whereas the fixed-arity variant of \textsc{louds} is easy to traverse in constant time per edge, the \textsc{dfuds} representation requires more space \cite{BDMRRR05,Nav16}: apart from the 4 bits, each node with $c$ children uses $c+1$ bits to mark its number of children.
%For instance, a node encoded as `\texttt{0101}' has children by symbols \texttt{0} and \texttt{3}. In the original \textsc{dfuds} one would need to store `\texttt{(()}', plus symbols \texttt{0} and \texttt{3} (using 2 bits each) to indicate which are the two children. This would require 7 bits to encode this node. 

As a consequence, our actual storage format cannot be traversed in constant 
time per edge. Rather, we will traverse the blocks sequentially and carry out
all the edge traversals or updates on the block in a single left-to-right
pass. This is not only cache-friendly, but convenient because we do not need
to store nor recompute any sublinear-space data structure to speed up traversals
\cite{Cla96}.

%As originally done for $k^2$-trees \cite{BLN13,BCPdBN17}, we choose not to store external nodes explicitly. In their case, however, there are no major complications when traversing the tree, mainly because the tree is represented in breadth-first and all external nodes have the same depth. However, this is not the case for a tree stored in depth-first: 
A complication related to our format is that, when traversing the tree, we must maintain the current trie depth in order to identify the leaves (these are always at depth $\log_2{n}$). Besides, as we traverse the block we must be aware of which are the frontier nodes, so as to skip them in the current block or switch to another block, depending on whether or not we want to enter into them.

%Finally, we decide to store no data structure for speeding-up traversals (as, for instance, \textsf{rank}, \textsf{select}, and balanced-parenthesis data structures). Although this can make traversals slower (as we must carry out sequential scans on the tree to navigate it), we save space and also time when inserting a node into a block (as there are no data structure using additional space and that must be rebuilt upon insertions). For instance, if we want to descend from node 0 to node 7 in the tree of Figure \ref{fig:k2tree}, we must traverse nodes 1 to 6 before getting to 7.

\paragraph{Operation $\mathsf{child}$.}
This is the main operation needed for traversing the tree. Let $\mathsf{child}(x, i)$ yield the child of node $x$ by symbol $0\le i \le 3$ (if it exists). Assume node $x$ belongs to block $B$. For computing $\mathsf{child}(x,i)$, we first check whether node $x$ is in the frontier of $B$ or not. To support this checking efficiently, we keep a finger $i_{\!f}$ on array $F_{\!B}$, such that $i_{\!f}$ is the smallest value for which $F_{\!B}[i_f]$ is greater or equal than the preorder of the current node in the traversal. Since we traverse in preorder, and $F_{\!B}$ is sorted, increasing $i_{\!f}$ as we traverse $T_{\!B}$ is enough to keep $i_{\!f}$ up to date. When the preorder of the current node exceeds $F_{\!B}[i_{\!f}]$, we increase $i_{\!f}$. If $F_{\!B}[i_{\!f}] = x$, then node $x$ is in the frontier, hence we go down to block $P_{\!B}[i_{\!f}]$, start from the root node (which is $x$ itself stored in the child block), and set $i_{\!f} \gets 0$. Otherwise, $x$ is not a frontier node, and we stay in $B$.

Determining whether the $i$-th child of a node $x$ exists requires a simple bit inspection. If it does, we must determine how many children of $x$ (and their subtrees) must be skipped to get to $\textsf{child}(x, i)$. We store a precomputed table that, for every 4-bit pattern and each $i=0,\ldots, 3$, indicates how many subtrees must be skipped to get the desired child. For instance, if $x$ is `\texttt{1011}' and $i = 2$, this table tells that one child of $x$ must be skipped to get to the node labeled 2. 

In our sequential traversal of $B$, corresponding to a depth-first traversal of
$T_{\!B}$, we keep a stack $S$ (initially empty) with the number of children 
not yet traversed of the ancestors of the current node.
We start looking for the desired child by moving to position $x+1$, corresponding to the first child of $x$ in preorder. At this point, we push the number of children of this node into $S$. The traversal is carried out by increasing an index on the array that stores $T_{\!B}$. The key for the traversal is to know where in the tree one is at each step. As said before, we keep track of the current depth $d$, to know when we arrive to frontier nodes. When traversing, we update $d$ as follows. Every time we move to the next node (in preorder), we increase $d$ only if (1) $d$ is not the maximum depth (minus 1, recall that the last level is not represented), (2) the current node is not a frontier node, or (3) the current node is the last child of its parent. We use $S$ to check the latter condition. Every time we reach a new node, we push in $S$ its number of children if the node is not of maximum depth (minus 1), and it is not a frontier node. Otherwise, we instead decrease the value at the top of the stack, since in both cases the subtree of the corresponding node has been completely traversed. When the top value becomes 0, it means that a whole subtree has been traversed. In such a case we pop $S$, decrease the current depth $d$, and decrease the new value at the top (if this also becomes 0, we keep repeating the process, decreasing $d$ and the top value). 

Once the stack $S$ becomes empty again, we have traversed the subtree of the first child. We repeat the same process from the current node, skipping as many children of $x$ as needed.

\paragraph{Operation $\mathsf{insert}$.}
To insert a point $(c, r)$, we use the corresponding Morton code $M= yz$, for strings $y\in\{0,\ldots,3\}^*$ and $z \in \{0,\ldots, 3\}^+$ to navigate the trie, until we cannot descend anymore. Assume that we have been able to get down to a node $x$ (stored in block $B$) that represents string $y$, and at this node we have failed to descend using the first symbol of $z$. Then, we must insert string $z$ in the subtree of node $x$. If the block has enough space for the $|z|$ new nodes, we simply find the insertion point from $x$ (skipping subtrees as explained above), make room for the new nodes, and write them sequentially using a precomputed table that translates a given symbol of $z$ to the 4-bit pattern corresponding to the unary node for that symbol. We also store a precomputed table that, given the encoding of $x$ and the first symbol of string $z$, yields the new encoding for $x$. 

If, on the other hand, the array used to store $T_{\!B}$ has no room for the new nodes, we proceed as follows. If the array is currently able to store up to $N_i < N_t$ nodes, we reallocate it to make it of size $N_j$, for the smallest $N_j$ such that $N_i + |z| \le N_j$ holds. If, otherwise, $N_i = N_{max}$, or $N_i + |z| > N_{max}$, we must first \emph{split} $B$ to make room. 

To minimize space usage, the splitting process should traverse $T_{\!B}$ to choose the node $w$ such that splitting $T_{\!B}$ at $w$ generates two trees whose size difference is minimum. We combine this criterion, however, with another 
one that 
optimizes traversal time. As explained, an advantage of our method is that we
can traverse several edges in a single left-to-right scan of the block. Such
scan, however, ends when we have to follow a pointer to another block.
We try, therefore, to have those pointers as early as possible in the block so
as to minimize the scan effort spent to reach them. Our splitting criterion, 
then, tries first to separate the leftmost node in the block whose subtree size
is 25\%--75\% of the total block size. 

After choosing node $w$, we carry out the split by generating two blocks, adding the corresponding pointer to the new child block, and adding $w$ as a frontier node (storing its preorder in $F_{\!B}$ and its pointer in $P_{\!B}$). 

\paragraph{Increasing the size of deeper blocks.}
A way to reduce the cost of traversing the blocks sequentially is to define a small maximum block size $N_{max}$. 
The cost is that this increases the space usage, because more blocks will be needed (thus increasing the number of pointers, and hence the space, of the data structure). 
We have the fortunate situation, however, that the most frequently traversed blocks are closer to the root, and these are relatively few. 
To exploit this fact, we define different maximum block sizes according to the depth of the corresponding block, with smaller maximum block sizes for smaller depths. 
We define parameters $0 \le d_1 < d_2$ such that blocks whose root has depth at most $d_1$ have maximum block size $N''_{max}$, blocks whose root has depth at most $d_2$ have maximum block size $N'_{max}$, 
and the remaining blocks have maximum size $N_{max}$, for $N''_{max} < N'_{max} < N_{max}$. In this way, we aim to reduce the traversal cost, while using little space at deeper blocks. 
Pushing this idea to the extreme, we may set $N''_{max}=1$, equivalent to allowing that the top part of the tree be represented with explicit pointers.

\paragraph{Analysis again.}
Theorem~\ref{thm:main} builds on a highly theoretical result \cite{ADR16}, thus
our engineered structure obtains higher time complexities. In our 
implementation, each operation costs $O(N_{max})$ time, which we set 
close to $\log^2 N$ to obtain the same space redundancies
of dynamic bitvectors. In turn, the
implementation of dynamic bitvectors \cite{BCPdBN17} takes $\Theta(\log^2 N)$ 
time per basic operation (edge traversal or update). An advantage of 
our implementation is that, during the $\Theta(\log^2 N)$-time traversal of a 
single block, we may process several $k^2$-tree edges, but this is not
guaranteed. As a result, we can expect that our implementation be about as 
fast as the dynamic bitvectors or significantly faster, depending on the tree 
topology. Our experiments in the next section confirm these expectations.

\section{Experiments}

\subsection{Experimental setup}

We experimentally evaluate our proposal comparing it with the dynamic
$k^2$-tree implementation based on dynamic bit vectors~\cite{BCPdBN17}, to demonstrate the comparative performance of our technique.
Other dynamic trie implementations exist~\cite{AS10,BBV10,K17} that
are designed for storing general string dictionaries, and could store the points
using their Morton codes. However, these techniques usually do not
compress and require space comparable to the original collection of strings;
moreover, even if they are more efficient to search for a single element, they
lack the ability to answer more complex queries, such as row/column queries, through a single traversal of the tree,
that is required in $k^2$-tree representations.

We use four different datasets in our experiments. Their basic information is described in
Table~\ref{tab:datasets}. The graphs \textsf{indochina} and \textsf{uk} are
Web graphs\footnote{\texttt{http://law.di.unimi.it/datasets.php}}, known to be very sparse and
compressible. The datasets \textsf{triples-med} and \textsf{triples-dense} are
selected predicates of the DBPedia 3.5.1\footnote{%
\texttt{https://wiki.dbpedia.org/services-resources/datasets/data-set-35/ data-set-351}},
transformed through vertical partitioning as in previous work~\cite{AGBFMPN14};
they are also sparse matrices but much less regular, and more difficult to
compress.

\begin{table*}[t]
\centering
\setlength{\tabcolsep}{10pt} % Default value: 6pt
\renewcommand{\arraystretch}{1}
\begin{tabular}{|l|l|r|r|}\hline
\multirow{2}{*}{Type} & \multirow{2}{*}{Dataset}         &   Rows/cols	& Points \\
					  &							         &  (millions)	& (millions) \\
\hline
\multirow{2}{*}{Web graph} & \textsf{indochina-2004} &   7.4 & 194.1 \\
							&
\textsf{uk-2002}    	&   18.5 & 298.1\\
							\hline
\multirow{2}{*}{RDF} & \textsf{triples-med} 		&   67.0 & 7.9  \\
					 &	\textsf{triples-dense} 	&   67.0 & 98.7 \\
\hline
\end{tabular}

\ \\
\caption{Datasets used in our experiments.}
\label{tab:datasets}
\end{table*}

Four our structure we use $k=2$ and the following configuration parameters: 
$N''_{max}=1$ (i.e., we use explicit pointers in the first few levels of the trie), 
$N'_{max}=96$, and use varying $N_{max}$, from 256 to 1024. 
We show the tradeoff using values of $d_1$ 8 and 12, and values of $d_2$ from 10 to 16 depending on $d_1$. 

For the
approach based on dynamic bitvectors (\texttt{dyn-bitmap}), we show results of the practical implementation with the default setup (block size 512 and
$k=4$ in the first 3 levels of decomposition) and, when relevant, another
configuration with smaller block size 128
and $k=4$ in the first 5 levels.

We run our experiments in a machine with 4 Intel i7-6500@2.5GHz cores and
8GB RAM, running Ubuntu 16.04.6. Our code is implemented in C++ and compiled with g++
5.5.0 using the -O9 optimization flag. 

\subsection{Results}

In order to test the compression and performance of our techniques, we start by building the representations from the original
datasets. To do this, we shuffle the points in the dataset into a random order, and insert them in the structures one by one. 
Then, we measure the average insertion time during construction of the
complete dataset, as well as the space used by the structure after
construction. 

\begin{figure*}[t]
   \centering
        \includegraphics[width=\textwidth]{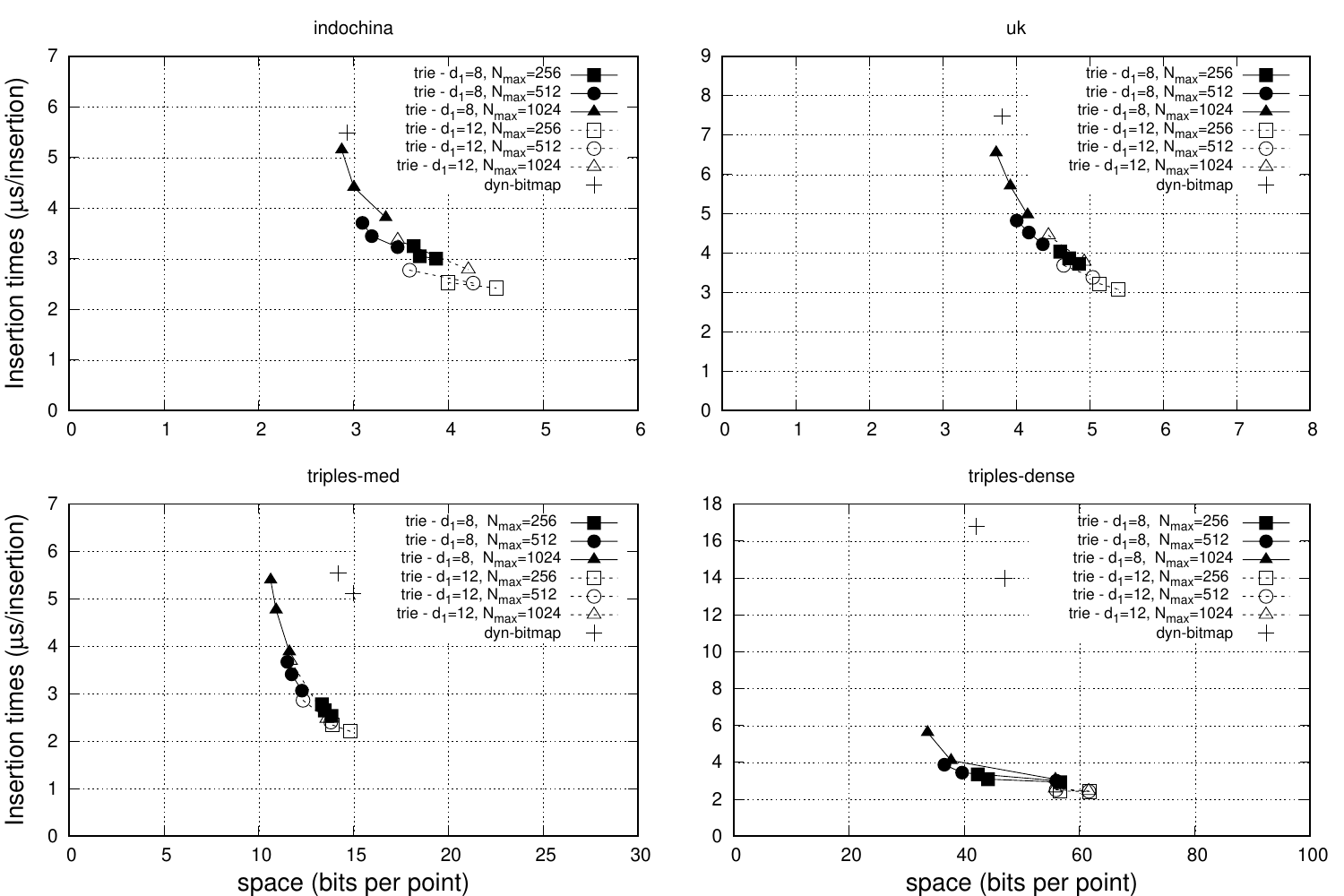}
    \caption{Compression and insertion times (in bits per inserted point and $\mu$s/insertion)}
    \label{fig:insert}
\end{figure*}

Figure~\ref{fig:insert} displays insertion times during construction and final
space for all the datasets and tested configurations. The results show that in Web
graphs (\textsf{indochina} and \textsf{uk}) our
representations can be created significantly faster than the dynamic bitvectors
while requiring negligible additional space, for example 20--25\% faster using 3\%
more space.
Moreover, our representations provide a wide space-time tradeoff that the
technique based on dynamic bitvectors does not match (in Web graphs we only
show results for the default configuration of \textsf{dyn-bitmap}, because
the configuration with smaller blocks is both larger and slower). The configuration to achieve this
tradeoff is also quite intuitive: larger(smaller) blocks in the lower levels lead to slower(faster), but more(less) compact structures.

In the RDF datasets (\textsf{triples-med} and \textsf{triples-dense}), our 
structures are even more competitive, using far less space and time
than the dynamic bitvectors. In \textsf{triples-med}, our structures
are 2.5 times faster when using similar space, or use 25\% less space for
the same speed. In \textsf{triples-dense} we are about 5 times faster when
using the same space, and still 3 times faster than dynamic bitvectors when
using 20\% less space.
Notice that the main difference between RDF and Web graph datasets
is the regularity and clusterization of the points in the matrix, which is much higher in Web graphs than in RDF datasets. This also explains the worse space results achieved in 
these datasets compared to Web graphs. A similar difference in regularity
exists between \textsf{triples-med} and \textsf{triples-dense}, where the latter is much more difficult to compress.

\begin{figure*}[t]
   \centering
        \includegraphics[width=\textwidth]{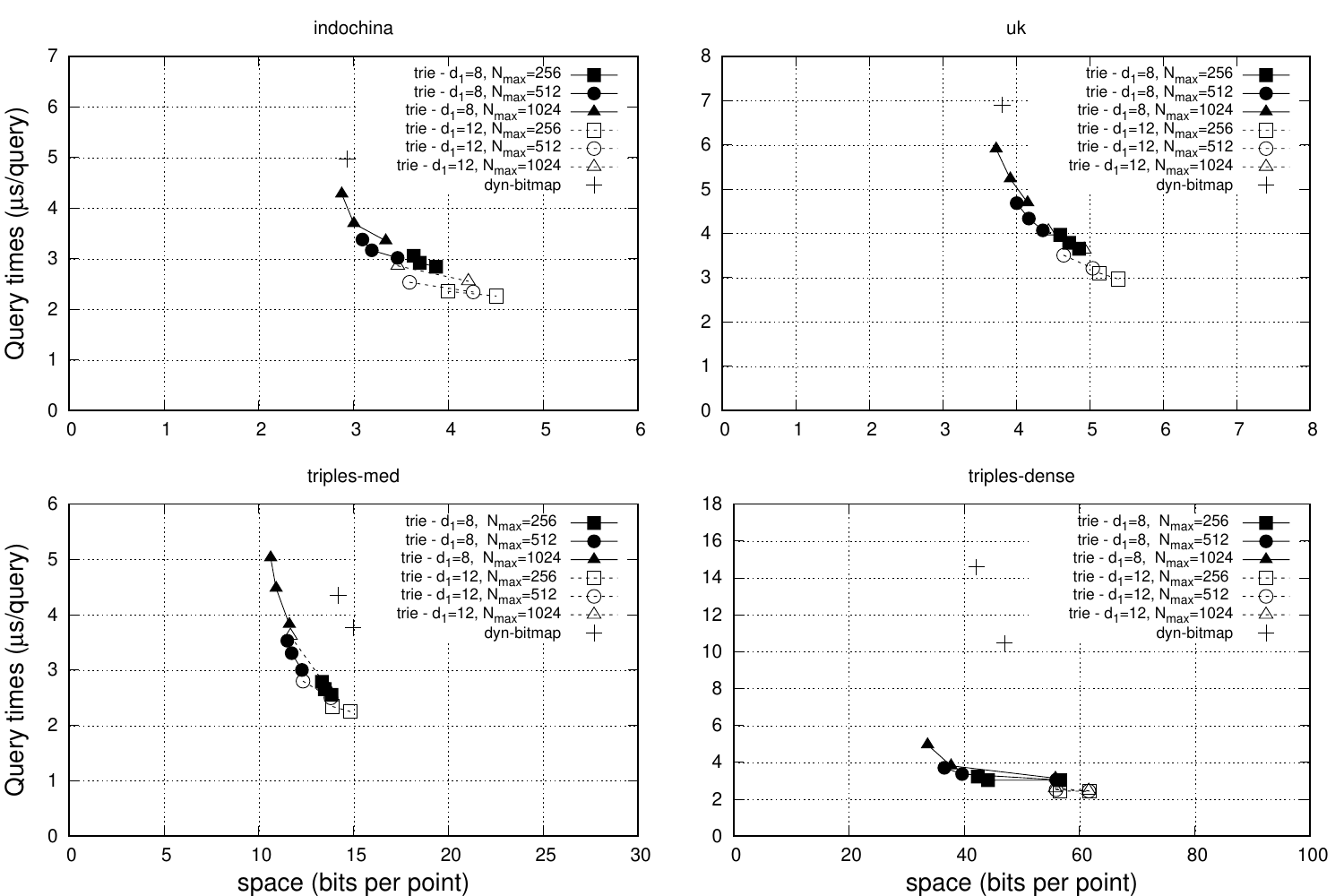}
    \caption{Query times to retrieve cells (in $\mu$s/query)}
    \label{fig:query}
\end{figure*}

Next, we measure the average query times to retrieve a point. To do this, we again select the points of each collection in random order, limiting our selection to 100 million points in the larger datasets,
and measure the average query time to search for each of them. Figure~\ref{fig:query} displays the query times for these cell retrieval queries. 
Results are analogous to those of insertion times. In Web graphs, our tries 
obtain even better performance compared to dynamic bitvectors.
In RDF datasets the times are slightly closer but our tries still outperform
dynamic bitvectors in space and time: In \textsf{triples-med} tries are 70\%
faster when using the same space, or 20\% smaller when taking the same time. 
In \textsf{triples-dense} tries are 4 times faster when using the same space,
and 3 times faster when using 20\% less space.

We also perform tests querying for 100 million randomly selected cells.
In practice, most of these cells will not belong to the collection, and they will probably be
relatively far from existing points, hence allowing the structures to stop the
traversal in the upper levels of the tree. These kind of queries are much faster and almost identical
for all the trie configurations tested in each dataset. 
In Web graphs, the dynamic bitvectors 
obtain better query times in Web graphs for these queries (0.4--0.6
$\mu$s/query in \textsf{indochina} and \textsf{uk}, while our tries take around
0.6--0.7 and 0.75--0.95 $\mu$s/query, respectively).
In RDF datasets, our tries are still significantly faster (around 0.55--0.6
$\mu$s/query in both datasets, whereas dynamic bitvectors take
1.1--1.2 $\mu$s/query in \textsf{triples-med} and 1.5--1.9 $\mu$s/query in
\textsf{triples-dense}). This points to the depth of the tree search as a relevant factor in query complexity:
our tries seem to have more stable query times, and are faster in queries that involve traversal of the full tree depth. In Web graphs, where points are usually clustered, non-existing points are detected in 
upper levels of the tree, and query times are usually better. In the RDF datasets, where points are more randomly distributed, the depth of the search is expected to be higher on average even if the dataset
is still very sparse.

\section{Conclusions}

Regarding the $k^2$-tree as a trie on the Morton codes of the points it 
represents yields a new view that differs from the classical one based on 
bitvectors
\cite{BLN13}. We have shown that this makes an important difference in the
dynamic scenario, because dynamic tries can break lower bounds on maintaining
dynamic bitvectors. Apart from the theoretical result, we have implemented a
dynamic trie specialized in representing $k^2$-trees, where the trie is cut
into a tree of blocks, each block representing a connected component of the
trie. The dynamic trie uses a depth-first search deployment of the trie, unlike
the classical level-wise deployment. While this format cannot be traversed in 
constant time per trie edge, it is convenient for a dynamic trie representation
because it is consistent with the tree of blocks, update operations require
local changes, a single left-to-right block scan processes several downward 
edge traversals, and such scan is 
cache-friendly and does not require rebuilding any speed-up data structure.

Our experimental results show that our representation significantly outperforms
the one based on dynamic bitvectors \cite{BCPdBN17} on some datasets, in space,
time, or both, depending on the nature of the dataset.

In the final version we will include experiments on other operations like 
extracting all the neighbors of a node. A future goal is to explore 
applications of our 
dynamic $k^2$-tree representation, in particular for graph databases 
\cite{AGFLP18}.

\bibliographystyle{splncs03}
\bibliography{paper}

\end{document}